\UseRawInputEncoding
\documentclass[twocolumn,aps,prl,superscriptaddress]{revtex4-1}
\usepackage[T1]{fontenc}

\usepackage{graphicx}
\usepackage{subfigure}
\usepackage{epsfig}
\usepackage{amssymb,amsmath,amsopn,amsfonts}
\usepackage{colordvi}
\usepackage{color}
\usepackage{subeqnarray}
\usepackage{bbold}
\usepackage{xcolor}
\usepackage{siunitx}
\usepackage{braket}
\usepackage[hidelinks]{hyperref}
\newcommand{\Appendix}{Supplementary Material}

\begin{document}

\title{On-chip quantum information processing with distinguishable photons}

\author{Patrick Yard}
\email{patrick.yard@bristol.ac.uk}
\affiliation{Quantum Engineering and Technology Laboratories, School of Physics and Department of Electrical and Electronic Engineering, University of Bristol, Bristol, UK}

\author{Alex E. Jones}
\email{a.jones@bristol.ac.uk}
\affiliation{Quantum Engineering and Technology Laboratories, School of Physics and Department of Electrical and Electronic Engineering, University of Bristol, Bristol, UK}

\author{Stefano Paesani}
\affiliation{Quantum Engineering and Technology Laboratories, School of Physics and Department of Electrical and Electronic Engineering, University of Bristol, Bristol, UK}
\affiliation{Center for Hybrid Quantum Networks (Hy-Q), Niels Bohr Institute, University of Copenhagen, Blegdamsvej 17, DK-2100 Copenhagen, Denmark}

\author{Alexandre Ma\"inos}
\affiliation{Quantum Engineering and Technology Laboratories, School of Physics and Department of Electrical and Electronic Engineering, University of Bristol, Bristol, UK}

\author{Jacob F. F. Bulmer}
\affiliation{Quantum Engineering and Technology Laboratories, School of Physics and Department of Electrical and Electronic Engineering, University of Bristol, Bristol, UK}

\author{Anthony Laing}
\affiliation{Quantum Engineering and Technology Laboratories, School of Physics and Department of Electrical and Electronic Engineering, University of Bristol, Bristol, UK}

\footnotetext[1]{To reach 99.9\% visibility interference for a 1\% surface code threshold~\cite{sparrow2017}, resonances must be aligned to within $\sim3\%$ of the linewidth.}

\date{\today}
\begin{abstract}

Multi-photon interference is at the heart of photonic quantum technologies.
Arrays of integrated cavities can support bright sources of single-photons with high purity and small footprint, but the inevitable spectral distinguishability between photons generated from non-identical cavities is an obstacle to scaling.
In principle, this problem can be alleviated by measuring photons with high timing resolution, which erases spectral information through the time-energy uncertainty relation.
Here, we experimentally demonstrate that detection can be implemented
with a temporal resolution sufficient to interfere photons detuned on the scales necessary for cavity-based integrated photon sources.
By increasing the effective timing resolution of the system from 200~ps to 20~ps,
we observe a $~20\%$ increase in the visibility of quantum interference between independent photons from integrated micro-ring resonator sources that are detuned by 6.8~GHz.
We go on to show how time-resolved detection of non-ideal photons can be used
to improve  the fidelity of an entangling operation
and to mitigate the reduction of computational complexity in boson sampling experiments.
These results pave the way for photonic quantum information processing with many photon sources without the need for active alignment.

\end{abstract}


\maketitle
\section{\label{sec:intro}Introduction}

Proposed photonic quantum technologies will require large numbers of photons \cite{wang2020integrated,bartolucci2021switch,li2015resource} and would benefit from cavity based single-photon sources. 
Integrated cavities can be used to increase the purity of parametric sources, while reducing their footprint and power consumption \cite{burridge2020high,llewellyn2020chip}, as well as increasing the generation rates from solid state sources based on two-level systems \cite{gazzano2013bright,somaschi2016near}.
However, fabrication imperfections produce an inherent misalignment of emission wavelengths from multiple cavity based sources.
While thermal \cite{silverstone2015qubit}, strain \cite{sun2013strain} or electrical \cite{akopian2010tuning, nowak2014deterministic, zhai2022} tuning techniques can be used to adjust the source emission wavelength post-fabrication,
scaling these techniques to many cavities is impractical; thermal and electrical cross-talk, for example, make aligning even small numbers of resonant sources to the required sub-GHz precision a challenge \cite{carolan2019scalable, llewellyn2020chip, paesani2019generation,Note1}.

Here, we experimentally address this challenge by demonstrating on-chip quantum interference of photons with distinguishable emission spectra generated from integrated cavity sources.  
Our approach, which does not use spectral filtering or active tuning, relies on fast photon detection to directly exploit the conjugate relationship between frequency and time:
provided photons are detected with a high enough timing resolution, their spectral information can be sufficiently erased to allow quantum interference between initially distinguishable photons \cite{legero2003time}.
Previous works have applied this technique with narrow-bandwidth cavity emission from atomic systems or parametric processes. Avalanche photodiodes were used to interfere photons with frequency detunings of tens of MHz  ~\cite{legero2004exp,wang2018experimental,zhao2014entangling}, which is insufficient for the GHz-scale detuning that is typical of integrated photon sources \cite{pfeifle2014coherent,zhai2022quantum}.
Here, we show that commercial superconducting nanowire single-photon detectors (SNSPDs) enable the erasure of spectral distinguishability at the GHz scale for photons generated in standard integrated micro-ring resonator (MRRs) sources.
We demonstrate the viability of this approach for photonic quantum information processing by performing a range of on-chip experiments using distinguishable photons, including time-resolved Hong-Ou-Mandel (HOM) interference, error-mitigated photonic fusion operations, and boson sampling experiments with up to three interfering (six detected) photons.
These results demonstrate that fast detectors can deliver powerful error mitigation for cavity-based sources, and can readily be extended to deterministic photon sources in heterogeneous photonic quantum information processors.

\begin{figure*}
    \centering
    \includegraphics[width = 1\linewidth]{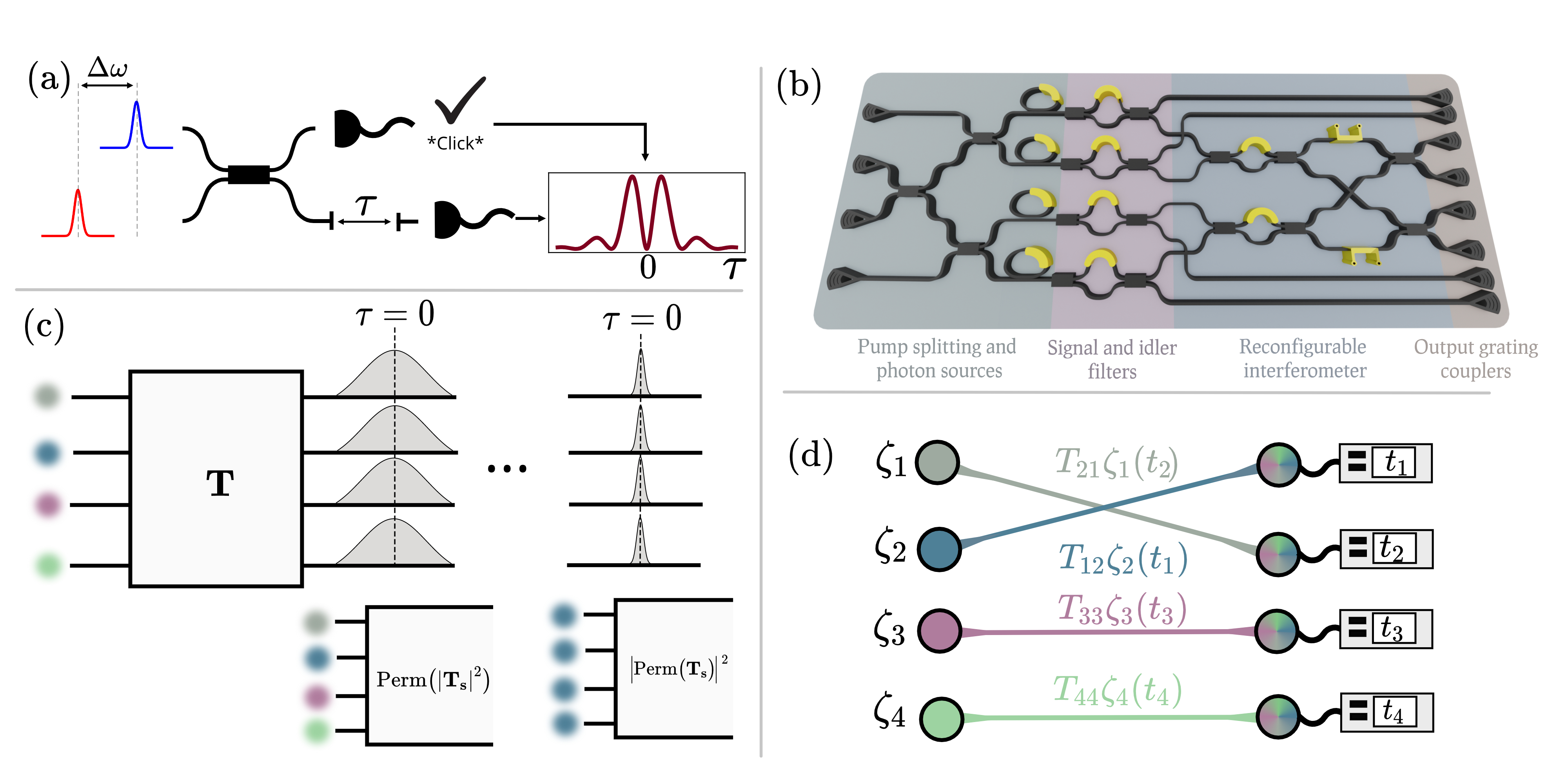}
    \caption{(a) \textbf{Time-resolved two-photon interference}. Distinguishable photons separated by frequency $\Delta\omega$ input to a balanced beam splitter lead to interference fringes in the time-domain. Detecting a photon in the first output projects the other photon into a superposition of coming from either of the inputs. This results in single-photon interference fringes in the second photon's relative arrival time $\tau$, with beat frequency $\Delta\omega$. (b) \textbf{Schematic of photonic chip}. Pulsed laser light enters the chip through grating couplers and pumps tuneable MRR sources, with thermo-optic phase shifters shown in gold. Tuneable filters separate idler and signal photons that are then interfered or used as heralds, respectively. Photons are detected off-chip by SNSPDs. (c) \textbf{Boson sampling using temporal filtering}. Low timing precision means no interference and output probabilities are given by permanents of positive real matrices. High timing precision and filtering around $\tau=0$ allow distinguishable photons to interfere, giving an output distribution described by permanents of the complex transfer matrices. (d) \textbf{Boson sampling from the distribution of photon arrival times}. Given input photons with temporal profiles $\zeta_i$, the output probability is given by the permanent of the transfer matrix weighted by the temporal amplitudes (Eq.~\ref{eq:multi_perm_main}). We depict one possible exchange of photons with its associated amplitude. The total probability is the squared magnitude of the sum of all possible amplitudes.}
    \label{fig:fig1}
\end{figure*}
\section{Time-resolved interference theory}
It has been shown for HOM interference that provided photons are detected exactly coincidentally at the output ports of a balanced beam splitter, bunching can be observed with photons of arbitrary and, in general, different spectro-temporal profiles \cite{legero2003time}.
Expressions for time-resolved interference statistics have since been extended to more photons and to include spectral impurity \cite{tamma2015multiboson,shchesnovich2020distinguishability}.
The probability of detecting $N$ photons with arrival times $\vec{t}=(t_{1},\ldots,t_{N})$ at the outputs of a linear optical interferometer described by scattering matrix elements $T_{ij}$ is (see Supplementary Material) 
\begin{equation}
    P_{\mathrm{coinc}}\left(\vec{t}\right) = \vert\mathrm{Perm}\left(\Lambda\right)\vert^2,
    \label{eq:multi_perm_main}
\end{equation}
where the matrix $\Lambda$ has elements given by $\Lambda_{ij} = T_{ij}\zeta_j(t_i)$ and $\zeta_i$ is the temporal profile of the $i^\mathrm{th}$ input photon. From this we see the complexity of quantum interference, arising from the calculation of permanents of complex matrices, is retained, even if the photons have different spectro-temporal profiles. 
This effect can be applied in the context of boson sampling \cite{tamma2016boson,tamma2016multi,laibacher2018toward} and also for 
entanglement swapping \cite{zhao2014entangling}. 
\begin{figure*}
    \centering
    \includegraphics[width = 1\linewidth]{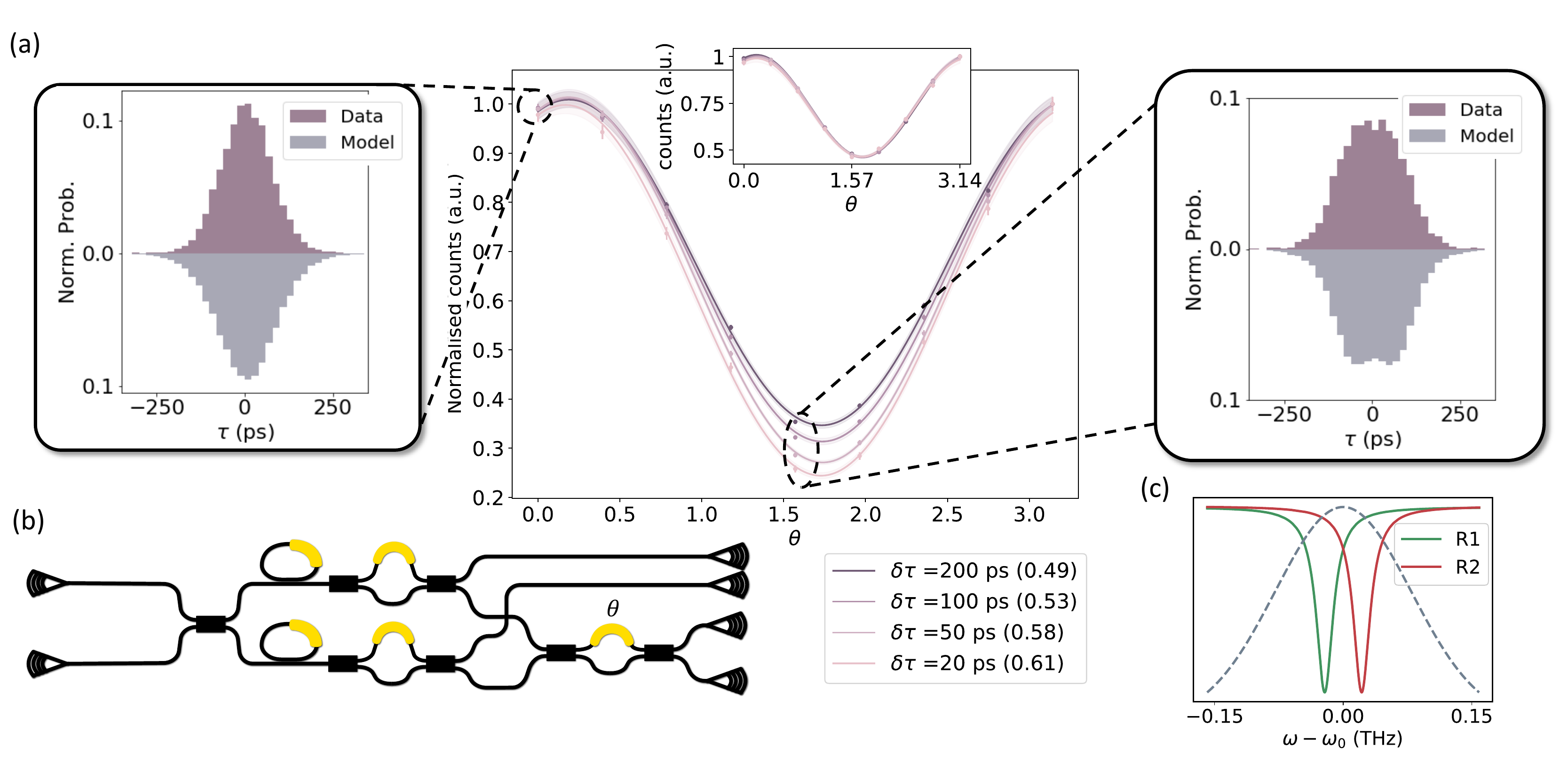}
    \caption{\textbf{Time-resolved HOM interference fringes}. (a) Central panel shows normalised coincidence counts as the MZI phase $\theta$ is varied, for different timing windows. The offset in the fringe minimum is due to crosstalk from AMZI filters. The visibility for each timing window is shown in the legend (below) and the shaded area around each line shows the 1-$\sigma$ interval, using the error from the fit. Poissonian error bars are smaller than the symbol size. Inset shows fringe for distinguishable photons. We also show histograms, at the peaks and dip, and compare to our model. (b) Schematic of subset of the chip used in this experiment, swept phase is also indicated. (c) Resonance positions of the two ring resonances relative to the pump pulse (shown as a dotted line)}
    \label{fig:fig2}
\end{figure*}
The complexity of Eq.\ref{eq:multi_perm_main} will be degraded by the imperfect timing resolution of detectors and time tagging electronics, characterised by the full width at half maximum (FWHM) of their response functions, often called the `jitter'.
We model the overall effect of detection equipment jitter by convolving interference fringes with a Gaussian distribution whose variance is the sum of the variances for the detectors and time-tagger channels used. 
\begin{figure*}
    \centering
    \includegraphics[width = 1\linewidth]{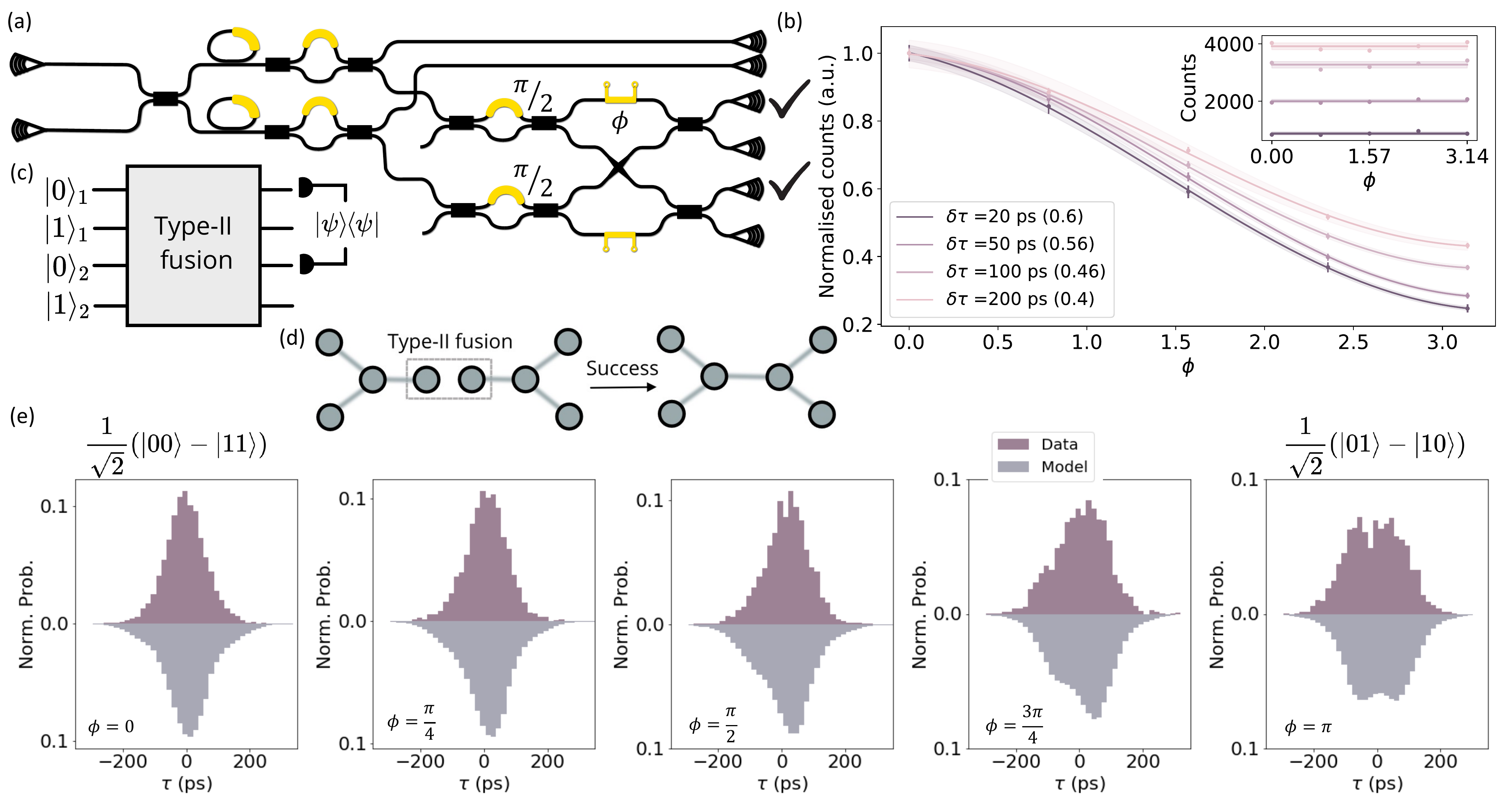}
    \caption{\textbf{Time-resolved interference and type-II fusions in a complex Hadamard}. (a) Chip configuration. Tick marks indicate the output pattern used in (b) and (e). (b) Fringes in $\phi$ for different timing windows showing that visibility increases with timing resolution. The inset shows the signal for temporally distinguishable photons generated by consecutive pump pulses: there is no dependence on $\phi$ or $\delta t$. (c) Type-II fusion gate on dual-rail photonic qubits, with qubit mode labels shown. The indicated output click pattern projects onto a Bell state $\ket{\psi} = \frac{1}{\sqrt{2}}(\ket{+-} + e^{i\phi}\ket{-+})$. (d) Effect of a type-II fusion gate on entangled resource states. (e) Measured relative arrival time histograms (purple above) and theoretical model (grey below) shown for each value of $\phi$.}
    \label{fig:fig3}
\end{figure*}

\section{Experimental setup}
A schematic of the integrated photonic device is shown in Fig.~\ref{fig:fig1}.
A pulsed pump laser is used to generate pairs of single photons through spontaneous four-wave mixing (SFWM). 
The laser is filtered and amplified before being passively split on-chip by a tree of multi-mode interference couplers, designed to implement balanced beam-splitters, and pumping up to four silicon micro-ring resonator sources. 
The spectra of these sources can be individually tuned by voltage-controlled thermo-optic phase shifters, and low-power continuous wave (CW) lasers are used to monitor and align the ring-resonators to the desired emission wavelengths.
Asymmetric Mach-Zehnder interferometer (AMZI) filters separate signal and idler wavelengths before the signals are coupled off-chip and the idlers enter a programmable integrated interferometer (see Fig.~\ref{fig:fig1}b). 

Photons are detected by high efficiency
and low jitter SNSPDs from PhotonSpot, with an average per-detector jitter measured to be $\sim$75~ps. Timing logic is performed by a Swabian Ultra II time-tagger, with single channel jitter of $\sim$15~ps.  
More detail on the experimental setup is given in the Supplementary Material.

\begin{figure*}
    \centering
    \includegraphics[width = 1\linewidth]{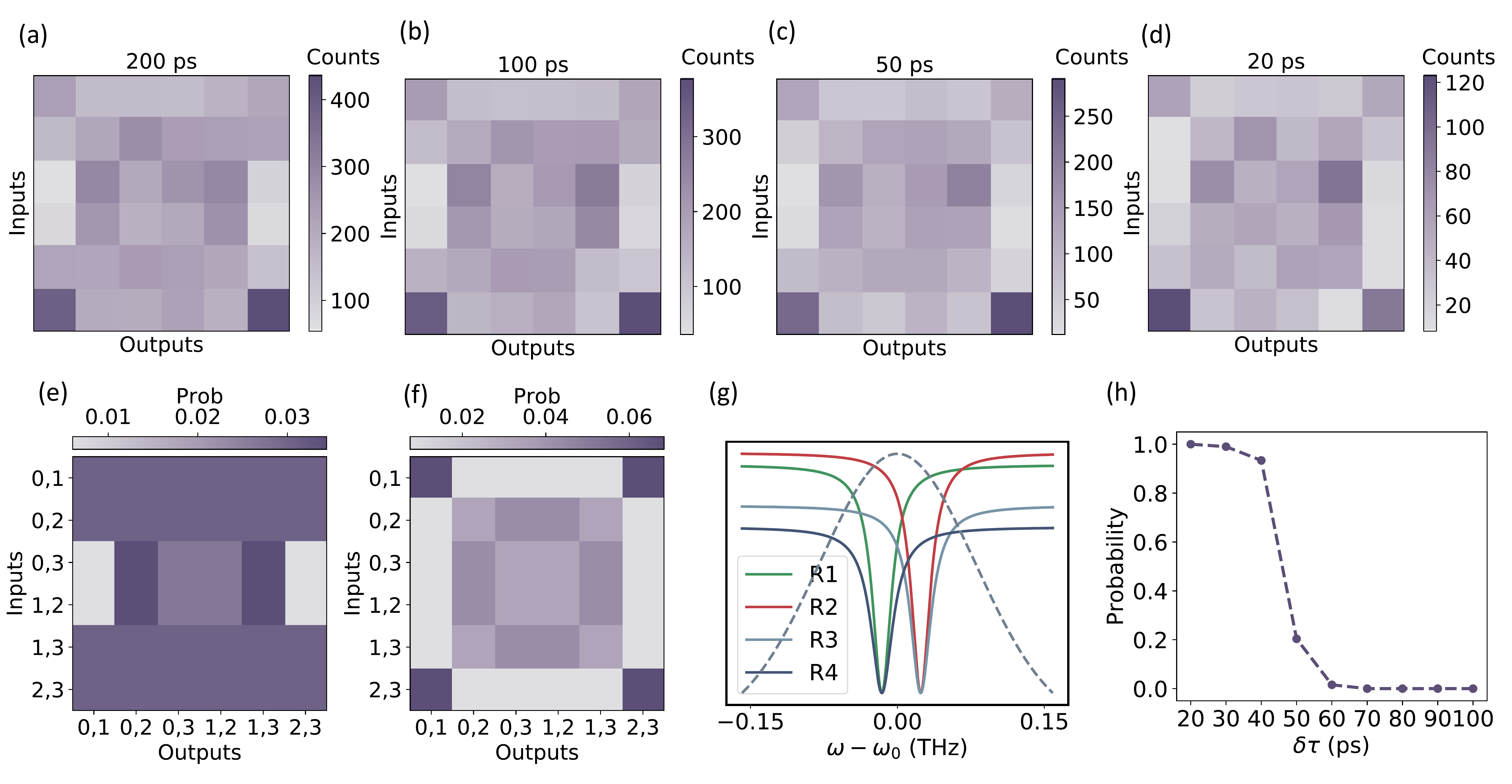}
    \caption{\textbf{Scattershot boson sampling with temporal filtering} (a)-(d) measured probability distributions for varying timing window. (e) Theoretical output distribution for the adversarial model. (f) Corresponding test model distribution. (g) Spectral alignment of the four ring resonator sources. Dotted line illustrates the spectral response of the pump pulse. (h) Final Bayesian probability for larger sample lists (2000 randomly chosen samples, averaged 300 times). For a timing window of 20~ps we use the entire sample list.}
    \label{fig:fig4}
\end{figure*}
\section{Time-resolved two-photon interference}

We first show that temporal filtering increases the interference visibility for two spectrally distinguishable photons, as depicted in Fig.~\ref{fig:fig1}c.
MRR resonances are detuned by 55~pm (6.8~GHz) and positioned symmetrically about the centre wavelength of the pump pulse (see Fig.~\ref{fig:fig2}c).
We then interfere the SFWM photons generated in these resonances on a Mach-Zehnder interferometer (MZI).
The output arrival times are recorded as the MZI phase $\theta$ is varied.
The relative arrival times of the interfering photons are calculated and binned into histograms with 20~ps bin width.
In Fig.~\ref{fig:fig2}a we plot the normalised MZI fringes for different effective timing resolutions, obtained by changing the timing window over which we integrate the depicted coincidence histograms.
The visibility of the resulting fringe in coincidence counts is defined as $V=(C_\mathrm{max}-C_\mathrm{min})/(C_\mathrm{max}+C_\mathrm{min})$, where $C_\mathrm{max}$ ($C_\mathrm{min}$) is the maximum (minimum) fringe counts within the chosen coincidence window.
The fringe visibility for non-interfering photons is $V=1/3$~\cite{vigliar2020}.
For a large integration window $\delta\tau=200$~ps, the MZI fringe visibility of $0.49\pm 0.01$ reflects some overlap of the detuned spectra shown in Fig.~\ref{fig:fig2}c.
As the integration window is narrowed, the number of counts decreases but the fringe visibility increases to a maximum of $0.61\pm 0.01$ when $\delta\tau=20$~ps, showing a significant improvement of the quantum interference via fast detection and temporal filtering. We find an average statistical fidelity between our model (see Supplementary Material) and the measured histograms of 0.984 $\pm$ 0.001.  The remaining distinguishability is due to the imperfect timing response of the detectors and time-tagging electronics used.
Moreover, in the inset figure we also show the fringes for temporally distinguishable photons generated by consecutive laser pulses (see Supplementary Material). 
These show no dependence on the timing window and all fringes have $V=1/3$, as expected.

\section{Fusion gates with time-resolved complex Hadamard interference}

We now demonstrate the use of time-resolved techniques to enhance a key building block for photonic quantum computing: the type-II fusion gate~\cite{browne2005resource}. Type-II fusion gates perform a partial Bell state measurement on two qubits, and can be used to fuse two resource states into a larger entanglement structure (see Fig.~\ref{fig:fig3}c,d). They play a central role in the generation of complex entangled structures for linear optical quantum computing \cite{bartolucci2021fusion}, as well as for generating entanglement in solid state systems \cite{barrett2005efficient} and in quantum repeater networks \cite{duan2001long,li2019experimental}. The photonic circuitry for type-II fusion gates can be performed by a four-mode complex Hadamard interferometer. This type of circuit, depicted in Fig.~\ref{fig:fig3}a, exhibits interesting multi-photon quantum interference features \cite{jones2020multiparticle,laing2012observation}, and finds extensive uses in classical encryption \cite{horadam2000cocyclic,hedayat1978hadamard}, dense coding \cite{werner2001all}, and the search for mutually unbiased bases (MUBs) \cite{durt2010mutually}.
We investigate how such features appear in a time-resolved complex Hadamard interference and analyse their use for fusion gates. 
For dimension $d=4$, all non-equivalent complex Hadamard matrices -- those that are not related by row and column permutations or phase multiplications -- are described by a single parameter $\phi$ in the family $F_4$ which, after normalisation, is given by
\begin{equation}\label{eq:f4}
    F_4\left(\phi\right) = \frac{1}{2} \begin{pmatrix}
       1&1 &1 & 1\\ 
       1& e^{i\phi}&-1 &-e^{i\phi} \\ 
       1& -1& 1& -1\\ 
       1&-e^{i\phi} &-1 & e^{i\phi}
       \end{pmatrix}.
\end{equation}
In Fig.~\ref{fig:fig3}a we show how, with input MZIs set to a balanced splitting ratio, our programmable integrated interferometer can implement this transfer matrix.
The output statistics for pairs of photons in certain input modes exhibit a dependence on the internal, controllable phase $\phi$.
This is because, while there are no closed loops inside the interferometer, there is interference of the pairs of paths that photons can take.
The time-resolved output coincidence probability is
\begin{equation}\label{eq:Pint_CH}
        P(t_0,t_0 + \tau) = \frac{1}{16}\vert\zeta_{1}(t_0 + \tau)\zeta_{2}(t_0) + e^{i\phi}\zeta_{1}(t_0)\zeta_{2}(t_0 + \tau)\vert^2.
 \end{equation}
When $\phi=\pi$ we recover the same expression as for a balanced beam splitter \cite{legero2003time}.
For other values, $\phi$ shifts the relative delays $\tau$ at which destructive interference suppresses coincidences.
This can be interpreted as adjusting the relative phase for conditioned single-photon interference, analogous to how applying a variable phase to a path in the double slit experiment shifts the fringe pattern on a screen. 
In Fig.~\ref{fig:fig3}e, we see how the complex Hadamard phase shifts the interference fringes as expected.
As before, we can integrate over the measured histograms to obtain an interference fringe. 
For indistinguishable photons this fringe should be given by $P_\mathrm{indist} = \frac{1}{2}\left(1 + \cos \phi\right)$ and for distinguishable photons there should be no dependence on $\phi$ \cite{jones2020multiparticle}. 
In Fig.~\ref{fig:fig3}b we see that as we change the timing integration window, we see an increase in the visibility of the fringe with decreasing timing window.
Figure.~\ref{fig:fig3}e shows good agreement between our model and measured histograms with an average statistical fidelity of 0.972 $\pm$ 0.002.

The above measurement probes the interference in the type-II fusion gate acting on a pair of photonic qubits initialized in the $\ket{00}$ input state. 
The value of $\phi$ has the effect of changing the Bell state that is projected onto the output click patterns. For example, the entangled state projected onto the output pattern shown in Fig.~\ref{fig:fig3}a,c is $\ket{\psi} =\left[ \left(1+e^{i\phi}\right)\ket{\Phi^{-}} + \left(1-e^{i\phi}\right)\ket{\Psi^{-}}\right]/\sqrt{2}$, where $\ket{\Phi^{\pm}}$ and $\ket{\Psi^{\pm}}$ are the canonical maximally entangled Bell states.
We can map the visibility of the fringe in Fig.~\ref{fig:fig3}b to the fidelity of the fusion gate as $\mathrm{F} = \frac{1+\gamma}{2}$ with $\gamma = \frac{2V}{1+V}$ \cite{rohde2006error}. 
An increase in the visibility of the fringe due to the temporal resolution thus corresponds to an improved fidelity of the fusion gate operation.
From the experimental fringe in Fig.~\ref{fig:fig3}b, we observe a significant improvement of the fusion gate fidelity, from 0.79 $\pm$ 0.02 to 0.875 $\pm$ 0.004, when tuning the timing window. 
Temporal filtering allows us to estimate the fusion gate fidelity, in practice, however, this results in a reduction of the gate success probability.
If, instead, a large timing window is used but the photon arrival times are measured precisely, then differences in the relative arrival times of the photons represent a random but heralded phase on the remaining photons and can, therefore, be accounted for with adaptive measurement \cite{campbell2007adaptive,zhao2014entangling}

\section{Boson sampling with spectrally distinguishable photons} 
\begin{figure*}
    \centering
    \includegraphics[width = 1\linewidth]{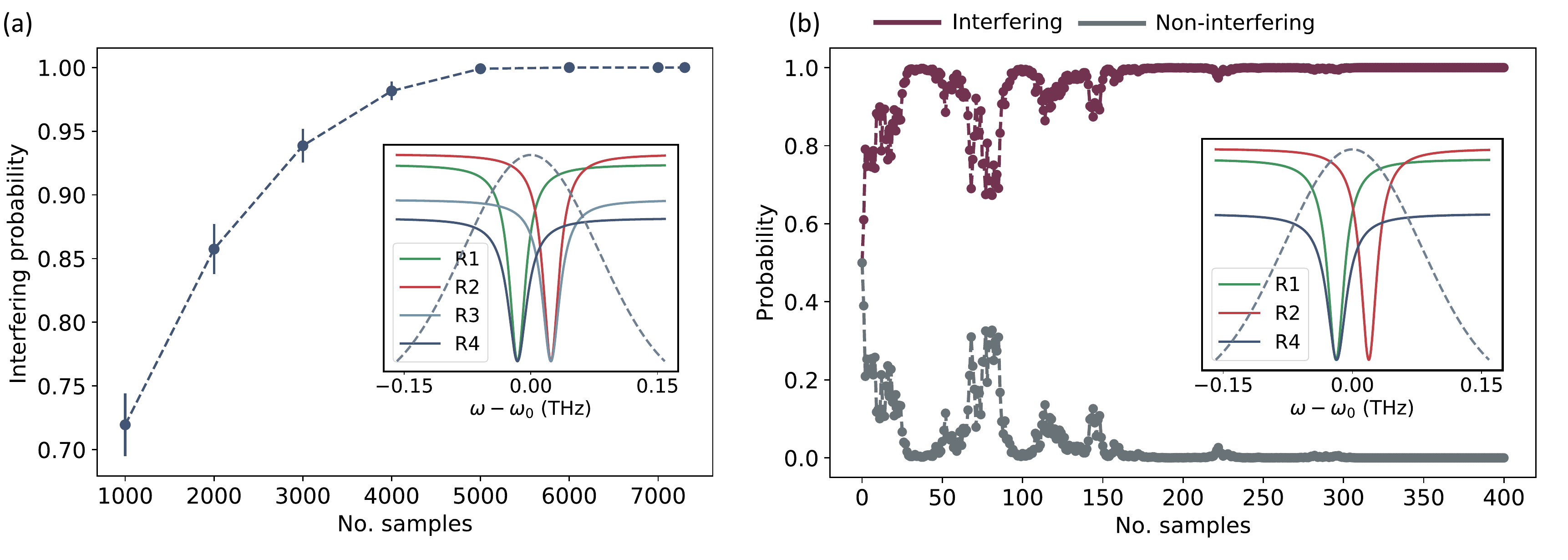}
    \caption{\textbf{Sampling from the distribution of photon arrival times} (a) Two photon scattershot boson sampling. Probability that a given sample string arose from our desired model as a function of sample size. Each sample string is drawn randomly from the total set of samples. We draw each sample size 200 times and plot the average and 1-$\sigma$ error bars. For the final point we take all measured samples. The inset shows the frequency alignment of the four ring sources. (b) Dynamic Bayesian updates showing the confidence that a given sample string arose from our test model.}
    \label{fig:fig5}
\end{figure*}
We now investigate the interference of multiple, spectrally distinguishable photons in boson sampling tasks.
Boson sampling is a model for quantum computation that, although not universal, can challenge the capabilities of classical computers, and has been demonstrated as a leading approach to demonstrating quantum advantage~\cite{aaronson2011computational,zhong2021phaseprogrammable,zhong2020quantum,madsen2022quantum}.
Multiple variants exist and all involve sampling photon numbers from a quantum state of light.
We first consider the scattershot variant~\cite{lund2014boson}: spontaneous sources are connected to the inputs of an interferometer and, when pumped simultaneously, a subset will generate photon pairs.
One photon from each pair heralds the input taken by its partner and so determines the scattering matrix for a particular run of the sampling experiment.
We use four MRR sources to perform two-photon scattershot boson sampling with distinguishable photons, and investigate using temporal filtering to recover the complexity of indistinguishable interference.
We align rings 1 and 4 to the same wavelength, and then rings 2 and 3 are aligned to another wavelength detuned by 54~pm (6.7~GHz), see Fig.\ref{fig:fig4}g.
The four-mode interferometer is programmed to implement $F_{4}(\pi/2)$ from Eq.~\ref{eq:f4}. 
We record the time tags for heralded two-photon events and, as earlier, can apply temporal filtering to coincidence histograms in post-processing.

We use Bayesian methods to determine whether our experimental samples were more likely drawn from a distribution involving significant quantum interference, or from one arising from probabilistic, distinguishable scattering \cite{paesani2019generation}.
We draw $N$ samples from our experiment, $x_{i}$, $i=1,\ldots,N$, such that the overall data $D=\{x_{i}\}$.
We then calculate the likelihood function
\begin{equation}
    P\left(M_{test}|D\right) = \frac{P\left(D|M_{test}\right)}{P\left(D|M_{test}\right) + P\left(D|M_{adv}\right)},
\label{eq:bayes}
\end{equation}
where $M_{test}$ is a test distribution that is hard to sample from, and $M_{adv}$ is an adversarial distribution that could describe the experiment but is easier to sample from.
A value of one indicates that the data $D$ were more likely drawn from $M_{test}$ than $M_{adv}$, and a value of zero indicates the opposite.
We assume samples are uncorrelated, so probabilities $P(D|M) = \prod_i^N P_{M}(x_i)$, where $P_{M}(x_{i})$ is the probability that sample $x_{i}$ came from distribution $M$.

In our case, $M_{test}$ comprises mainly indistinguishable interference involving permanents of complex matrices $\sim\vert \mathrm{Perm}(T_{s})\vert^{2}$, but also a small contribution from distinguishable scattering depending on permanents of real matrices $\sim\mathrm{Perm}\vert T_{s}\vert^{2}$.
This is to counter sensitivity of the likelihood test to the observation of photon detection patterns that are ideally suppressed by the Fourier interferometer employed, but can happen in our experiment due to imperfections (see Supplementary Material for details). 
The adversarial model $M_{adv}$ consists of distinguishable scattering for the input patterns where the photons are spectrally distinguishable, and of indistinguishable interference for the input patterns where the photons are spectrally indistinguishable.

In Figs.~\ref{fig:fig4}a--d, we show the measured output distribution as a function of timing window. 
These plots include corrections for the input squeezing and output coupling losses, as well as subtraction of double emissions. 
We see a qualitative change, moving from the adversarial distribution in Fig.~\ref{fig:fig4}e to the test distribution in Fig.~\ref{fig:fig4}f.
Fig.~\ref{fig:fig4}h shows the average final Bayesian probability for sets of 2000 samples, chosen uniformly from the total sample list, as a function of the timing window used. 
As we decrease the timing window, we see a transition in which model best describes the data at between 30 and 60~ps, indicating recovery of quantum interference despite distinguishability.

Rather than filtering and incurring loss, it is instead possible to use the precise photon arrival times at the outputs as part of the sample from a boson sampler.
Similarly to the random but known input ports in the scattershot approach, the arrival times indicate random but known complex factors that alter the effective scattering matrix (as depicted in Fig.~\ref{fig:fig1}d).
This approach was introduced by Tamma et al.~\cite{tamma2015multiboson} and has been shown to be at least as hard as conventional boson sampling \cite{tamma2016boson,tamma2016multi,laibacher2015physics,laibacher2018toward}.
We experimentally perform two-photon scattershot and three-photon standard boson sampling where we explicitly include the photons' arrival times.
For the two-photon scattershot, we use the same resonance positions and interferometer as for the temporal filtering experiment.
But for the three-photon experiment, we only pump three MRRs and reduce the detuning to 5.9~GHz to ensure sufficient squeezing for six-fold events.
We also program a different interferometer as no phase-dependent three-photon interference appears in $F_{4}$ and both indistinguishable and distinguishable distributions are uniform for all input and output combinations (see Supplementary Material).

To verify our experiment we, again, use the Bayesian likelihood function, given in Eq.~\ref{eq:bayes}.
The ideal case is now given by Eq.~\ref{eq:multi_perm_main}, with the corresponding transfer matrix determined by the measured input and output pattern. 
We choose a non-interfering adversarial model with $P_{\mathrm{adv}}\left(\vec{t}\right) = \mathrm{Perm}\left(|\Lambda|^2\right)$. 
As above, we include a non-interfering term in the test model.
For two-photon, time-resolved scattershot, we show the averaged Bayesian probability for our test model for multiple sets containing a given number of samples in Fig.~\ref{fig:fig5}a.
For the three-photon case, due to the lower total number of events, we show the cumulative sample-by-sample probability that the data was drawn from the desired distribution. 
This dynamically updated confidence is shown in Fig.~\ref{fig:fig5}b, showing that our samples were more likely drawn from the interfering distribution.

\section{Discussion}

We have demonstrated photonic quantum information processing experiments with MRRs detuned up to 6.8~GHz and with photon linewidths of $\sim$3.8~GHz, which are both two orders of magnitude higher than previous experiments.
MRRs are promising candidates to produce high photon numbers \cite{arrazola2021quantum} but experiments have been limited to small numbers of sources due to the experimental challenge of aligning many sources \cite{carolan2019scalable}. 
To perform experiments with no active alignment of MRRs, we require that the maximum detuning, given by half the resonator free spectral range, is less than the maximum possible frequency erasure.
In this work, the per-detector jitter, was measured to be $\sim$75~ps, which is already sufficient to observe quantum interference between photons detuned by approximately twice the photon linewidth, fulfilling this condition.
Further analysis (see Supplementary Material) shows that using
the lowest jitter currently available from commercial SNSPDs ($15$~ps) \cite{singlequantumEOS},
photons detuned up to \SI{190}{\giga \hertz}
can result in  HOM visibilities above the quantum limit,
while photons detuned up to \SI{5.1}{\giga \hertz}
can result in visibilities above $\sim95\%$, which is sufficient to demonstrate quantum advantage \cite{renema2018efficient}. 
Using detectors with the lowest reported jitter in the literature (3~ps) \cite{korzh2020demonstration}, we could expect to observe 
non-classical interference from photons detuned by up to \SI{300}{\giga \hertz},
and visibilities above the quantum advantage threshold for photons detuned by \SI{29}{\giga \hertz}. 
This work shows that time-resolved detection can mitigate the demand to actively align resonant sources and significantly reduce device complexity for large scale quantum photonics experiments.

\section{Acknowledgements}

We acknowledge support from the Engineering and Physical Sciences Research Council (EPSRC) Hub in Quantum Computing and Simulation (EP/T001062/1). Fellowship support from EPSRC is acknowledged by A.L. (EP/N003470/1). S.P. acknowledges funding from the Cisco University Research Program Fund nr. 2021-234494. P.Y. would like to thank Emma Foley, Brian Flynn and Naomi Solomons for useful discussions. 
\bibliography{timeres_combined}
\bibliographystyle{apsrev4-1}

\clearpage
\pagebreak

\widetext
\begin{center}
	\textbf{\large \Appendix}\\ \vspace{9pt}
\end{center}

\setcounter{equation}{0}
\setcounter{figure}{0}
\setcounter{table}{0}
\setcounter{page}{1}
\makeatletter
\renewcommand{\theequation}{A\arabic{equation}}
\renewcommand{\thefigure}{S\arabic{figure}}
\renewcommand\@biblabel[1]{[#1]}
\renewcommand{\citenumfont}[1]{#1}
\makeatother

\subsection{N-photon time-resolved interference}\label{supp:general}

In the following analysis we look to derive the probability of detecting $N$ photons at $N$ different times $\vec{t} = (t_1, t_2, \dots, t_N)$ at the output of an $M$ mode linear optical interferometer, given input photons with arbitrary temporal spectra, $\zeta_i(t)$. For simplicity we assume that our photons are injected into the first $N$ input modes, and measured in the first $N$ output modes. 
Therefore, our measurements at times given by $\vec{t}$ project onto the state
\begin{equation}
    \ket{\psi_\text{coinc}(\vec{t})} = \prod_{i=1}^N \hat{b}^\dagger_i (t_i) \ket{0}^{\otimes M}
\end{equation}
where $\hat{b}^\dagger_i (t_i)$ is the creation operator acting on mode $i$ at time $t_i$.
The input state is given by
\begin{equation}
    \ket{\psi_\text{in}} = \prod_{i=1}^N \left( \int dt \zeta_i(t) \hat{a}^\dagger_i (t) \right) \ket{0}^{\otimes M}
\end{equation}
which after evolving through an interferometer becomes
\begin{equation}
    \ket{\psi_\text{out}} = \prod_{i=1}^N \left( \int dt \zeta_i(t) \left( \sum_{j=1}^M T_{ji} \hat{b}^\dagger_j (t) \right) \right) \ket{0}^{\otimes M}.
\end{equation}
Here, the $T_{ji}$ are the elements of the interferometer scattering matrix $\mathbf{T}$. 
We can now calculate the coincidence probability as $P_\text{coinc}(\vec{t}) = \left|\braket{\psi_\text{coinc}(\vec{t}) | \psi_\text{out}}\right|^2 $.
By inspecting the overlap between $\ket{\psi_\text{out}}$ and $\ket{\psi_\text{coinc}}$, we can see that the only the terms which will survive from $\ket{\psi_\text{out}}$ are those which have exactly one creation operator per mode for the first $N$ modes, and none in the other modes.
This also lets us remove the integrals, as for example
\begin{equation}
    \int dt \bra{0} \hat{b}_k(t_k) \zeta_i(t) T_{ji} \hat{b}^\dagger_j(t) \ket{0} = \delta_{kj} \zeta(t_j) T_{ji}.
\end{equation}
As shown, for example in Ref.~\cite{scheel2004permanents}, we can therefore write the final probability as 
\begin{align}
    P_\text{coinc} & =  \left| \sum_{\sigma \in S_N} \prod_{i=1}^N T_{i,\sigma(i)} \zeta_{\sigma(i)} (t_i) \right|^2 \\
    & = |\mathrm{Perm}(\Lambda)|^2,
\end{align}
where $S_N$ is the group of $N-$dimensional permutations and $\Lambda$ has elements given by $\Lambda_{ij} = T_{ij}\zeta_j(t_i)$. We can see that the amplitude for this state is the permanent of the matrix $\Lambda$ which is formed from the scattering matrix, weighted by the temporal shape of the photons. 
Here, we note that we have recovered the result of Ref.~\cite{tamma2015multiboson}. 

\subsection{Time-resolved HOM interference with mixed input states}\label{supp:mixed}

To determine the effect of impure input photons, we detail the derivation of the coincidence probability of time-resolved HOM interference with photons in spectrally mixed states. To start we define an arbitrary photon density operator to be 
\begin{equation}
    \rho_i = \sum_k \lambda_k \ket{\phi_{ik}}\bra{\phi_{ik}},
\end{equation}
where
\begin{equation}
    \ket{\phi_{ik}} = \int dt \; \zeta_{ik}(t) \hat{a}^\dagger_i(t) \ket{0}_i.
\end{equation}
We can then write the two photon input state $\ket{11}$ as 
\begin{equation}
    \rho_\mathrm{in} = \rho_1 \otimes  \rho_2 = \sum_{k l} \lambda_k \lambda^\prime_l \ket{\psi_{kl}}\bra{\psi_{kl}},
\end{equation}
with 
\begin{equation}
    \ket{\psi_{kl}} = \iint dt dt^\prime \;\zeta_{1k}(t)\zeta_{2l}(t^\prime) \hat{a}^\dagger_1(t)\hat{a}^\dagger_2(t^\prime)\ket{00}.
\end{equation}
Applying the beam-splitter relations to $\rho_{in}$ 
\begin{equation}
    \begin{aligned}
    \ket{\psi_{kl}} &\rightarrow \ket{\psi_{kl}}_\mathrm{out} = \iint dt dt^\prime \;\zeta_{1k}(t)\zeta_{2l}(t^\prime) \left[\hat{a}^\dagger_3(t) + \hat{a}^\dagger_4(t)\right]\left[\hat{a}^\dagger_3(t^\prime) - \hat{a}^\dagger_4(t^\prime)\right]\ket{00} \\
    &=  \iint dt dt^\prime \;\zeta_{1k}(t)\zeta_{2l}(t^\prime) \left[\hat{a}^\dagger_3(t)\hat{a}^\dagger_3(t^\prime) + \hat{a}^\dagger_4(t)\hat{a}^\dagger_3(t^\prime) - \hat{a}^\dagger_3(t)\hat{a}^\dagger_4(t^\prime)-\hat{a}^\dagger_4(t)\hat{a}^\dagger_4(t^\prime)\right]\ket{00},
    \end{aligned}\label{eq:psi_out}
\end{equation}
to give a final density matrix
\begin{equation}
    \rho_\mathrm{out} = \sum_{k l} \lambda_k \lambda^\prime_l \ket{\psi_{kl}}_\mathrm{out} \bra{\psi_{kl}}_\mathrm{out}.
\end{equation}
Using the time-resolved coincidence projector, $\hat{P}\left(t_1,t_2\right) = \hat{a}^\dagger_3(t_1) \hat{a}^\dagger_4(t_2)\ket{0}\bra{0}\hat{a}_4(t_2)\hat{a}_3(t_1)$, we arrive at 
\begin{equation}
    \begin{aligned}
        P_\mathrm{coinc}\left(t_1,t_2\right) &= \mathrm{Tr}\left[\rho_\mathrm{out}\times\hat{P}\left(t_1,t_2\right)\right]\\
        &=  \sum_{k l} \lambda_k \lambda^\prime_l\; {}_\mathrm{out}\bra{\psi_{kl}} \hat{P}\left(t_1,t_2\right) \ket{\psi_{kl}}_\mathrm{out} \\
        &= \sum_{k l} \lambda_k \lambda^\prime_l \vert\bra{0}\hat{a}_4(t_2)\hat{a}_3(t_1) \ket{\psi_{kl}}_\mathrm{out} \vert^2 \\
        &= \sum_{k l} \lambda_k \lambda^\prime_l \bigg|\iint dt dt^\prime \zeta_{1k}(t)\zeta_{2l}(t^\prime)\\& \qquad \qquad \times \big[\bra{0}\hat{a}_4(t_2)\hat{a}_3(t_1)\hat{a}^\dagger_4(t)\hat{a}^\dagger_3(t^\prime)\ket{0} -\bra{0}\hat{a}_4(t_2)\hat{a}_3(t_1) \hat{a}^\dagger_3(t)\hat{a}^\dagger_4(t^\prime) \ket{0}\big]\bigg|^2
    \end{aligned}
\end{equation}
To simplify this further, we note that $\bra{0}\hat{a}_4(t_2)\hat{a}_3(t_1)\hat{a}^\dagger_4(t^\prime_2)\hat{a}^\dagger_3(t_1^\prime)\ket{0} = \delta_{t_1t^\prime_1}\delta_{t_2t^\prime_2}$. This leaves us with only two non-zero terms from the double integral, giving a final probability of 
\begin{equation}
    P_\mathrm{coinc}\left(t_1,t_2\right) = \sum_{k l} \lambda_k \lambda^\prime_l \big| \zeta_{1k}(t_2)\zeta_{2l}(t_1) -  \zeta_{1k}(t_1)\zeta_{2l}(t_2)\big|^2.
\end{equation}
The final probability is therefore given by an incoherent sum of pure state probabilities which for $t_1 = t_2$ returns the indistinguishable probability, regardless of the number or specific shapes of the Schmidt modes. For other timing combinations, even with arbitrarily high timing precision, the incoherent sum masks the quantum interference. We can generalise this result to $N$ photon interference and arbitrary scattering matrices
\begin{equation}\label{eq:final_prob_mixed}
    P(\vec{t}) = \sum_{k_1 \cdots k_N} \prod^N_{i=1} \lambda_{i,k_i}\bigg\vert\mathrm{Perm}\left(\Lambda^{k_1 \cdots k_N}\right)\bigg\vert^2,
\end{equation}
where the matrix $\Lambda^{k_1 \cdots k_N}$ has elements $\Lambda^{k_1 \cdots k_N}_{ij} = T_{ij}\zeta_{j,k_j}\left(t_i\right)$, $k_i$ represents the $k^\mathrm{th}$ Schmidt mode of the $i^\mathrm{th}$ input photon, and $\zeta_{i,k_i}$ is the corresponding temporal profile. 
\subsection{Experimental setup}\label{sec:exp}
\begin{figure}[h!]
    \centering
    \includegraphics[width = 1\textwidth]{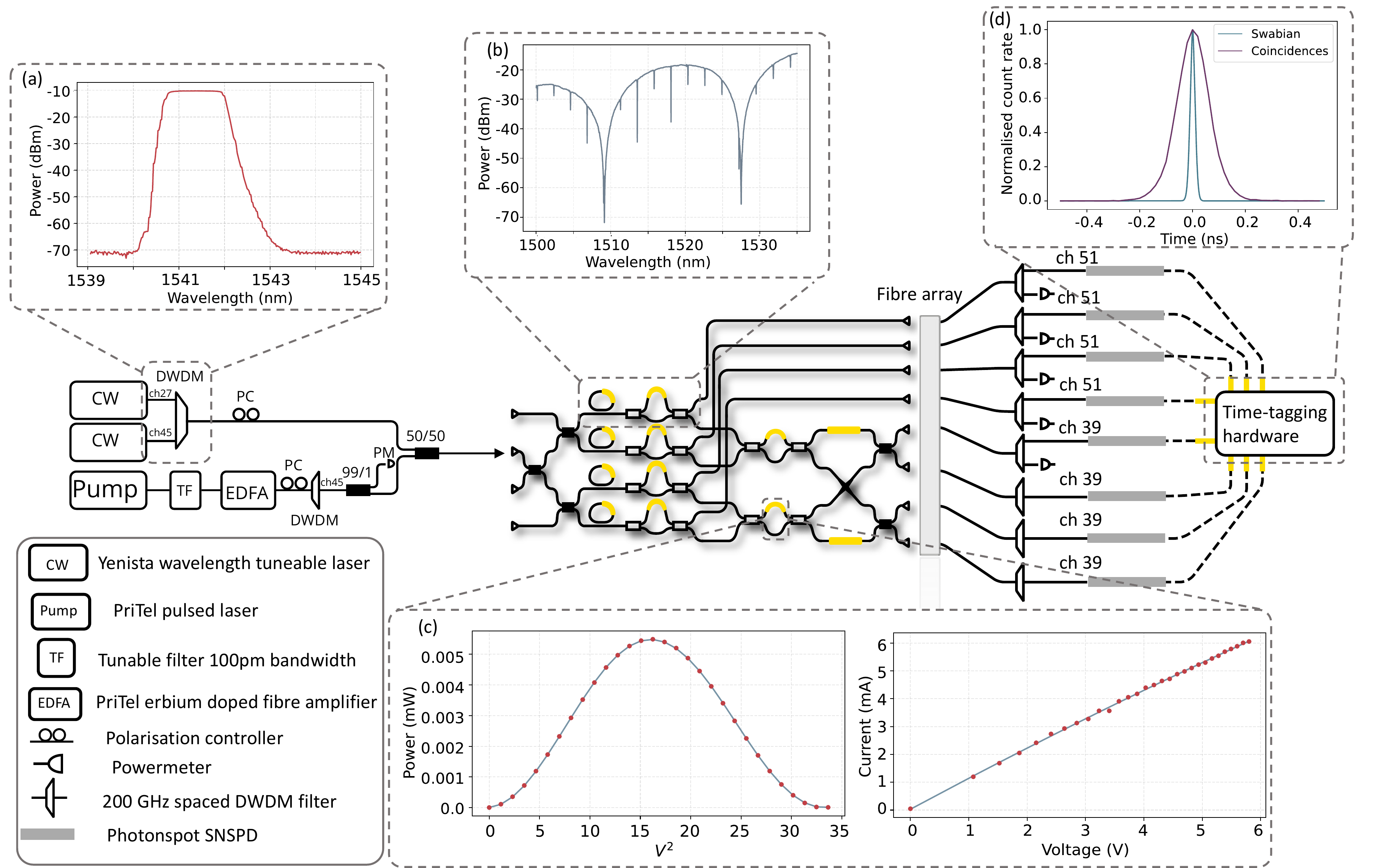}
    \caption{Detailed experimental schematic along with characterisation of important components. (a) Typical spectral response of a DWDM channel. (b) Spectral response at one of the heralding outputs. Combines the ring and AMZI spectra. (c) Characterisation of a thermo-optic heater. Left panel shows the transmission of an MZI as a function of applied voltage. Right panel shows the current-voltage relationship for one heater.(d) Typical coincidence histogram from one ring source. Also included is the response when using the internal test signals on two time-tagger channels. This is used to determine the intrinsic time-tagger jitter.}
    \label{fig:chip_charac}
\end{figure}

A detailed schematic of the experimental setup is shown in Fig.~\ref{fig:chip_charac}.
One high-powered continuous wave laser from Yenista is used for characterisation of optical components as well as aligning the ring resonator photon sources to a primary wavelength. 
This is used to align the pump resonances of the rings together.
For the experiments where we require aligning of ring sources to two different wavelengths we employ a second CW Yenista laser. 
To ensure that each ring is aligned to the desired wavelength, this second laser is used to align on resonances $\approx$\SI{9.5}{\nano \meter} away from the pump, this is chosen such that this wavelength is also transmitted correctly by the on-chip AMZI filters.
For generating photons, we employ pulsed pump lasers from PriTel.
For the initial two photon interference experiments we use a PriTel wavelength tuneable laser with a repetition rate of \SI{50}{\mega \hertz} and for the boson sampling experiments we use a PriTel ultrafast optical clock with a repetition rate of \SI{500}{\mega \hertz}. 
Both emit pulses with a \SI{2}{\nano \meter} bandwidth, however we use a tuneable filter to reduce the bandwidth to \SI{100}{\pico \meter} before amplification by a PriTel erbium doped fibre amplifier (EDFA).
A DWDM is then used to filter any potential unwanted side-band generation from the amplification.
The CW lasers are combined using a multichannel DWDM filter before being combined with the pump laser on a 50/50 beam-splitter.
Polarisation controllers on both CW and pump arms are used to maximise fibre to chip coupling. 
The on-chip component uses ring resonator photon sources (measured parameters given in table~\ref{tab:ring_params}) to generate two-mode squeezed vacuum states which are split on chip using AMZI filters which have free spectral ranges (FSR) of \SI{20}{\nano \meter}. 
\begin{table}[h!]
    \centering
\begin{tabular}{|c|c|c|c|}
 \hline
  Ring   & FSR (nm)& $\Delta\lambda$ (pm) & ER (dB) \\
  \hline
   R1  & 2.27$\pm$0.01 &29.14$\pm$1.13& 3.45$\pm$0.02 \\
   R2  & 2.27$\pm$0.01&29.12$\pm$1.09 &3.73$\pm$0.02\\
   R3  & 2.27$\pm$0.01&32.88$\pm$1.05& 7.46$\pm$0.03\\
   R4  & 2.27$\pm$0.01&34.45$\pm$1.04& 12.14$\pm$0.05\\
   \hline
\end{tabular}
    \caption[Ring resonator parameters]{Table showing the measured ring parameters for the 4 sources employed in this work. FSR is the free spectral range of the resonator. $\Delta\lambda$ is the resonance line-width and ER is the extinction ratio. } \label{tab:ring_params}
\end{table}
Signal photons (\SI{1536.36}{\nano\meter}, ITU channel 51) are coupled directly off-chip using grating couplers.
The idler photons (\SI{1546.1}{\nano\meter}, ITU channel 39) enter an integrated interferometer before being coupled off-chip. 
DWDM filters are used to reject the pump and send the single photons to PhotonSpot superconducting nanowire single photon detectors (SNSPD).
Powermeters are used on the output of a 99/1 beamsplitter and each herald output to monitor the input pump power and align the ring sources, respectively.
Coincidence logic is handled by a Swabian Ultra, used to record coincidences directly as well as saving raw time tags. 
To characterise the on-chip thermo-optic phase shifters we first measure the current-voltage curve of the heater, fitting with a second order polynomial to account for non-Ohmic effects.
This allows us to accurately map a particular voltage to the dissipated power.
To map the power to a phase, we sweep an interference fringe on the desired heater. 
The fit of this fringe, along with the IV fit, allows us to convert the voltage set into a phase.
To determine the system jitter of the experiment we can directly measure the Swabian jitter using the low-jitter internal test signal.
Using this signal on two channels allows us to measure a coincidence histogram with the FWHM dominated by the combined electronic jitter of both channels.
Assuming, that both channels have the same jitter, we find that the the single channel FWHM is 15.736 $\pm$ 0.002~ps.
To determine the combined jitter of detectors and time-tagging channels, we measure a coincidence histogram between photons generated through SFWM in a spiral waveguide, with a CW pump.
The detector jitter is much larger than the correlation time between the generated photons and we can therefore take the coincidence histogram to be the convolution of the detector and Swabian channels. 
Modelling each of these jitters as a Gaussian, and assuming that the detectors each have the same jitter, we find the single detector channel FWHM jitter to be 75.4 $\pm$ 1.6~ps.
\begin{figure}[h!]
    \centering
    \includegraphics[width = 1\linewidth]{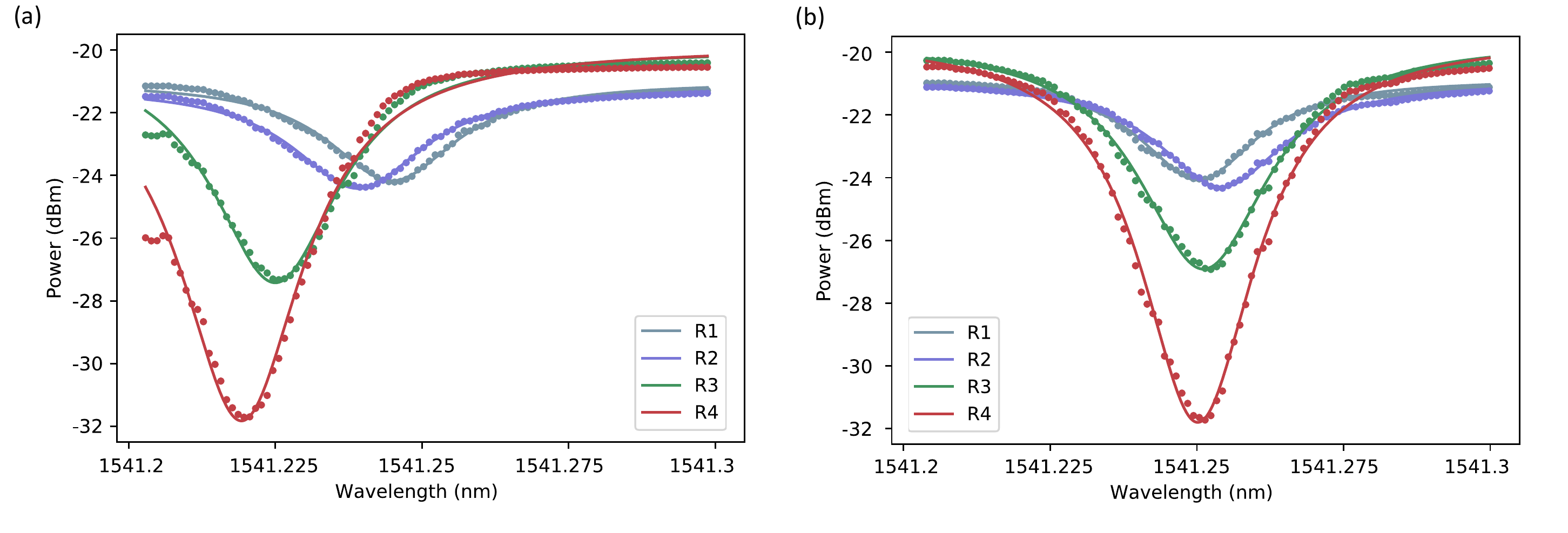}
    \caption[Effect of alignment algorithm]{Effect of optimisation protocol for ring alignment. Alignment wavelength set to \SI{1541.25}{\nano \meter}. (a) Ring resonances without optimisation. Here each ring is set separately by setting the voltage that minimises the transmitted power. Standard deviation of ring positions is \SI{10.5}{\pico \meter} (b) Rings aligned with Nelder-Mead optimisation algorithm. Starting voltages selected from Gaussian distributions centred at the voltages used in (a). Standard deviation of ring positions is \SI{1.6}{\pico \meter}.}
    \label{fig:locking_effect}
\end{figure}
\subsection{Aligning ring resonances}\label{supp:locking}

The ability to tune the resonance position thermally also means that thermal fluctuations in the lab can cause the ring resonances to drift.
To combat this the chip rig is attached to a Peltier and a PID temperature controller to keep the temperature of the chip stable. 
We also enclose the chip rig in a box in order to minimise changes in air temperature from affecting the chip. 
To combat cross-talk between thermo-optic phase shifters we use an optimisation loop to align the ring resonances to the desired wavelength \cite{carolan2019scalable}.
By fixing the CW laser at the frequency we wish to align to, we can define a cost function to be the sum of all powers transmitted by the rings we wish to align. 
Using the ring heater voltages as the system parameters an optimisation loop based on the Nelder-Mead algorithm is used to minimise the cost function.
The starting voltages were sampled from a Gaussian distribution centred at the voltage found by minimising the transmitted power for each ring individually.
The standard deviation of the distribution was chosen to be wide enough to allow restarts to escape any local minima but narrow enough to make sure the algorithm would converge.
In Fig.~\ref{fig:locking_effect}, here we compare the situation where we align each ring to the CW laser independently, by simply choosing the voltage value that minimises the light transmitted through each ring individually, and the case where we use the alignment mechanism described above.

\subsection{Data taking and background subtraction}\label{sec:data_taking}
\begin{figure}[h!]
    \centering
    \includegraphics[width = 1\textwidth]{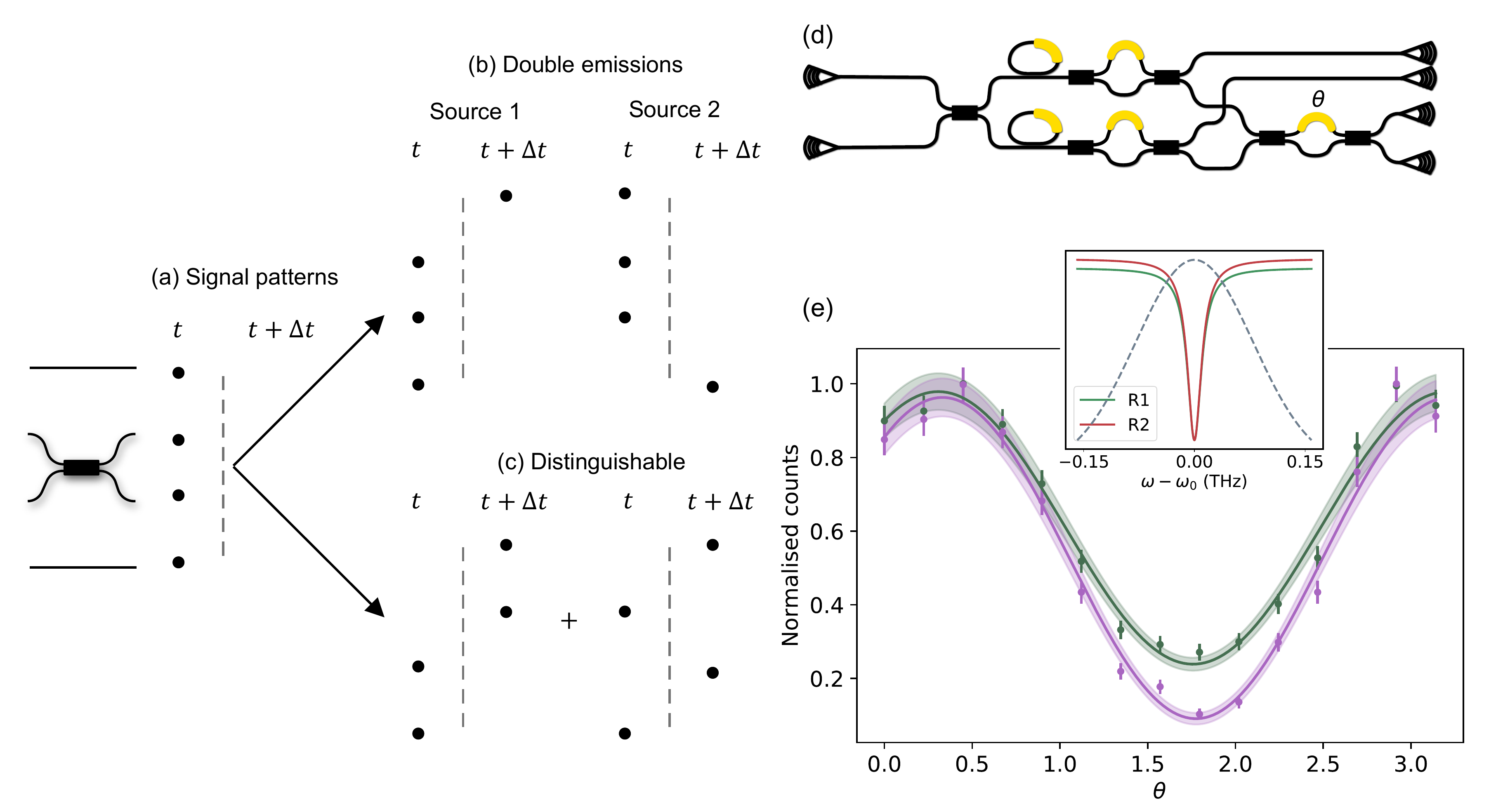}
    \caption[Virtual channels concept]{Use of virtual channels in order to measure background and distinguishable events. $t$ represents the original channel - i.e. photons generated in one pulse. $t + \Delta t$ represents a copy of a given channel but with all time tags delayed by $\Delta t$, when $\Delta t$ is set to the repetition rate of the laser, we now compare photons generated in the subsequent pulse.  (a) All photons generated in the same pulse. These correspond to the interference event we wish to observe. (b) Double emissions. One herald delayed, three photons generated in one pulse - i.e. a double emission from the source whose herald isn't delayed. (c) One pair of photons delayed. Corresponds to interfering photons being generated in subsequent pulses and therefore are distinguishable in time. (d) Chip configuration for heralded HOM interference with indistinguishable photons. Fringe acquired by sweeping angle $\theta$ and measuring four-fold coincidences (two herald and two signal) (e) Measured fringe with background (green) and without background (purple) including Poissonian error bars. Shaded area shows the $1\sigma$ interval from the errors on the fitting parameters. Visibility of green curve with no background subtraction $0.608 \pm 0.007$. With background subtraction this increases to $0.828 \pm 0.019$. Inset shows alignment of ring resonators relative to the pump pulse. }
    \label{fig:virtual_channels}
\end{figure}
We use the Swabian time tagger to save the time tags of all coincidence events which we can then post process as desired.
We also employ the ability of the Swabian time tagger to create copies of channels at a hardware level.
These virtual channels can be delayed relative to the original.
If the channels are delayed by a multiple of the repetition rate of the laser we can measure coincidences from photons generated in different pulses. 
By changing the pattern of virtual channels we use we can access the double emissions from each ring and also the events with no interference. 
Figure \ref{fig:virtual_channels} illustrates the different patterns which are relevant for two photon interference.
Figure \ref{fig:virtual_channels}a shows the case where all four detected photons are created in the same pulse, the main interference signal used in this work. 
The double emissions of each ring are found by shifting one herald channel, as shown in  Fig. \ref{fig:virtual_channels}b.
This can be generalised to more than one source by sequentially delaying each herald channel. 
The total of all these events consists of $N-1$ times the double emission rate.
We also use the case where interfering photons are generated in separate pulses and are therefore distinguishable in time and do not interfere.
This is shown in Fig. \ref{fig:virtual_channels}c.

\subsection{Time-resolved fringes}\label{supp:fringes}

The linear spectral response of a ring resonator should be a function $l(\omega,\omega_0,\delta \omega)$, such that $|l(\omega,\omega_0,\delta \omega)|^2$ gives us a Lorentzian function peaked at $\omega_0$ with a FWHM $\delta \omega$. From Ref~\cite{vernon2017truly} this is given by
\begin{equation}
    l(\omega,\omega_0,\delta \omega) = \sqrt{\frac{\delta \omega}{2\pi}}\times\frac{1}{-i\left(\omega - \omega_0\right) +\frac{\delta \omega}{2}}.\label{eq:ring_spectrum}
\end{equation}
The temporal profile, given by the Fourier transform of the spectrum, is a decaying exponential with a decay constant that is inversely proportional to resonance width, $\delta \omega$.
\begin{figure}[h!]
    \centering
    \includegraphics[width = 1\textwidth]{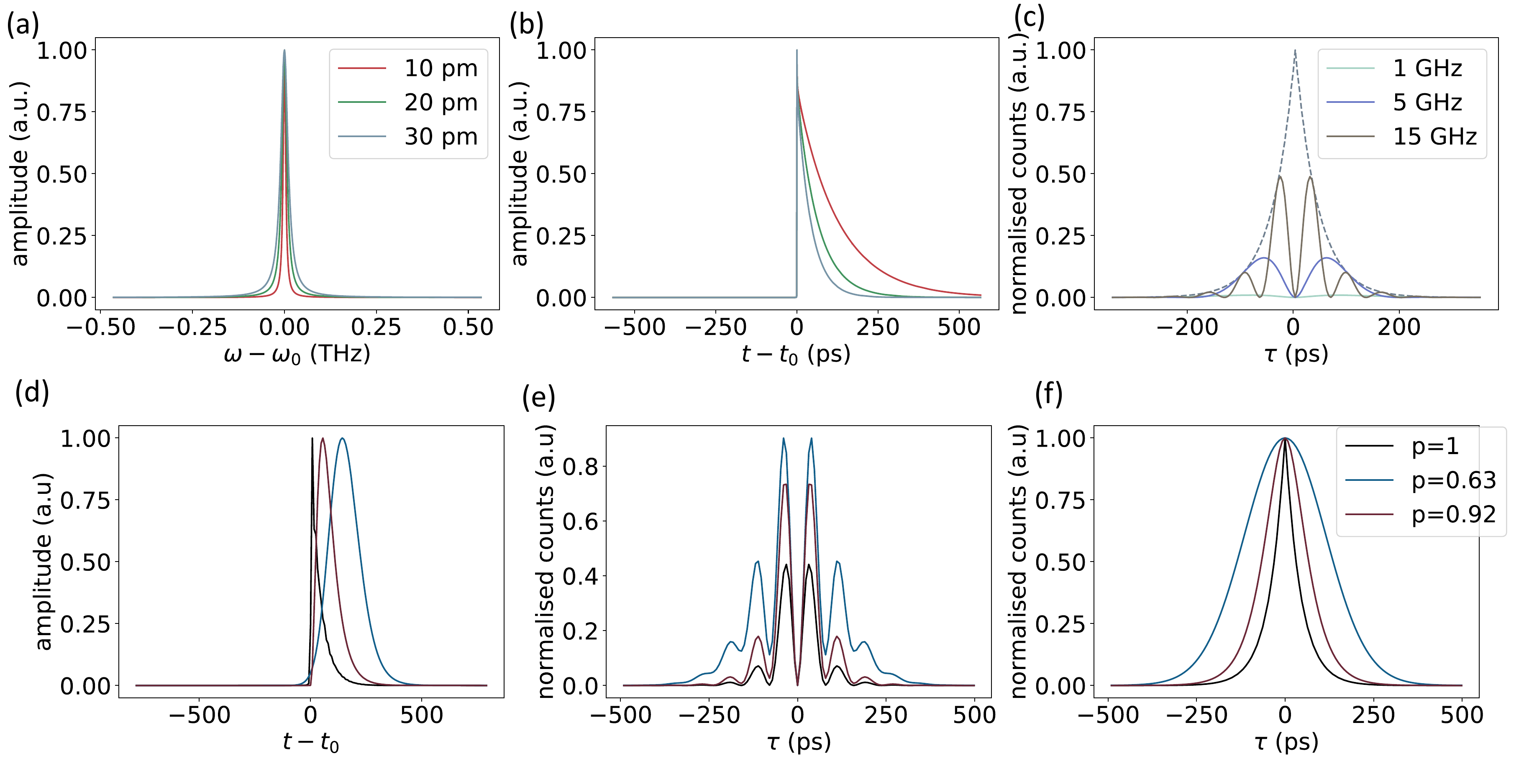}
    \caption{Two-photon interference simulations (a) Spectral profile of a ring resonance for different spectral widths. (b) Corresponding temporal profiles for (a). Time-resolved fringes as a function of temporal delay between photons for different detunings of the photon's central frequency. Fringes beat at the frequency of the detuning. (d) Total temporal profile acquired by incoherently summing the pure Schmidt modes. Black line shows temporal profile of a pure state. (e) Time-resolved fringes for a detuning of \SI{12.6}{\giga\hertz} for different photon purities. (f) Modulating envelope for different photon purities.}
    \label{fig:time_res_sims}
\end{figure}
To simulate the mixed photons arising from resonant sources, we calculate the joint spectral amplitude as
    %
        \begin{equation}
            F(\omega_s,\omega_i) = \int d\omega_p\alpha(\omega_p) \alpha(\omega_s + \omega_i-\omega_p)l(\omega_p) l(\omega_s + \omega_i-\omega_p) \phi(\omega_p,\omega_s,\omega_i)l(\omega_s)l(\omega_i),
        \end{equation}
        %
where $\alpha(\omega)$ is the spectral amplitude of the pump and the phasematching condition $\phi(\omega_p,\omega_s,\omega_i) \approx 1$ $\forall \omega_p,\omega_s,\omega_i$.
The Fourier transform of this function gives the joint temporal amplitude (JTA) $T(t_1,t_2)$.
The temporal Schmidt modes can then be calculated via a singular value decomposition of the JTA. 

As we wish to find fringes in the relative delay between detected photons we express the array of detection times in Eq.~\ref{eq:final_prob_mixed} as $\vec{t} = \{t_0,t_0+\tau_1, \cdots , t_0 + \tau_{N-1}\}$.
Integrating over $t_0$ results in interference fringes that depend on $N-1$ relative delays.
We illustrate these fringes for two photons on a balanced beam-splitter in Fig.~\ref{fig:time_res_sims}c for different detunings of the photons central frequency.
We see sinusoidal fringes, which beat at the frequency detuning of the photons.
These fringes are modulated by an envelope function which for ring resonator photons is a Lorentzian shape. 
In Fig.~\ref{fig:time_res_sims}d - e we show the effect of reducing the purity of the input photons.
We tune the photon purity by changing the width of the pump function. A narrower pump pulse results in lower purity photons. 

\subsection{Future projections}\label{supp:projections}

To analyse the scalability of this technique we use our model to determine HOM fringe visibilities as a function of photon detuning for a variety of likely system jitters. We fix the both photon line-widths to be \SI{30}{\pico \meter}. Total system jitter results from the convolution of two detector and two time tagger channels. Assuming Gaussian responses, and that both detector and time tagger channels have the same jitter, the total FWHM is given by
\begin{equation}
    \mathrm{FWHM}_\mathrm{tot} = \sqrt{ 2\mathrm{FWHM}_\mathrm{det} + 2\mathrm{FWHM}_\mathrm{tag}}.
\end{equation}
We examine three jitter regimes. First we use detector and time tagger jitter used in this experiment, corresponding to a total jitter FWHM of \SI{109}{\pico \second}. Secondly, we use the best commercially available numbers. Detector jitter taken to be \SI{15}{\pico \second} \cite{singlequantumEOS} and time tagger jitter to be \SI{4.71}{\pico \second} \cite{swabianx_datasheet}. Finally we use numbers corresponding to the state of the art, \SI{3}{\pico \second} detector jitter \cite{korzh2020demonstration} and assume this is the only source of jitter.
\begin{figure}[h!]
    \centering
    \includegraphics[width = 0.5\textwidth]{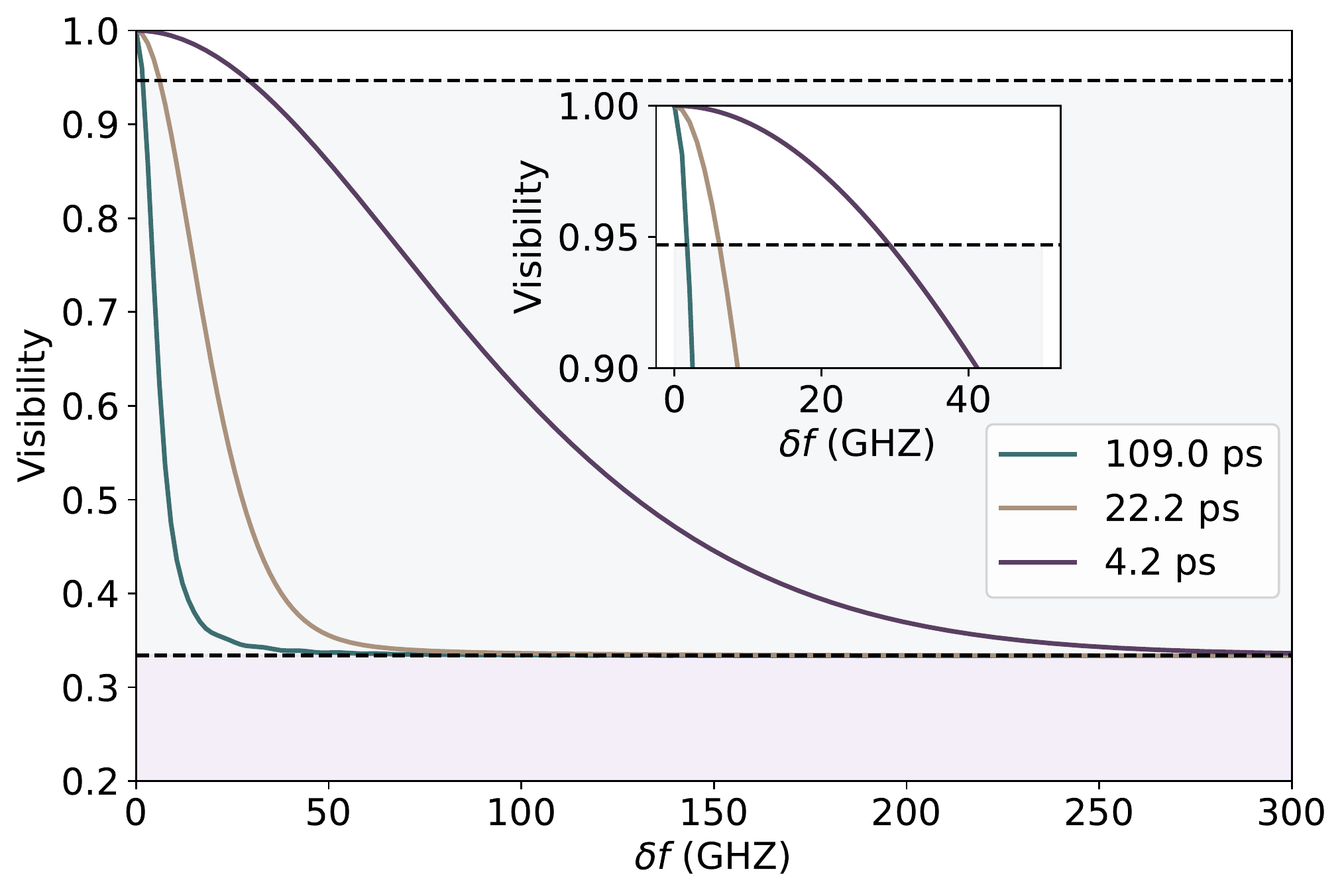}
    \caption{Projected HOM fringe visibilities as a function of photon detuning. We consider three different system jitters. The same as measured in this experiment, corresponding to a total system jitter of \SI{109}{\pico \second}, the best available commercial equipment corresponding to a total system jitter of \SI{22.2}{\pico \second} and the state of the art SNSPD jitter, corresponding to a total system jitter of \SI{4.2}{\pico \second}}
    \label{fig:future_proj}
\end{figure}
%
%


\subsection{Boson sampling with indistinguishable photons}\label{supp:BS}

To verify that our experimental setup is sufficient for boson sampling we perform two-photon scattershot and three-photon standard boson sampling with indistinguishable photons. We avoid three-photon scattershot as the loss in our system means that photons generated from the fourth source would become prevalent and would add noise to the data. For two photons, the four fold rate is high enough that we can reduce the squeezing in the sources and this noise process is less dominant. 
\begin{figure}[h!]
    \centering
    \includegraphics[width = 1\textwidth]{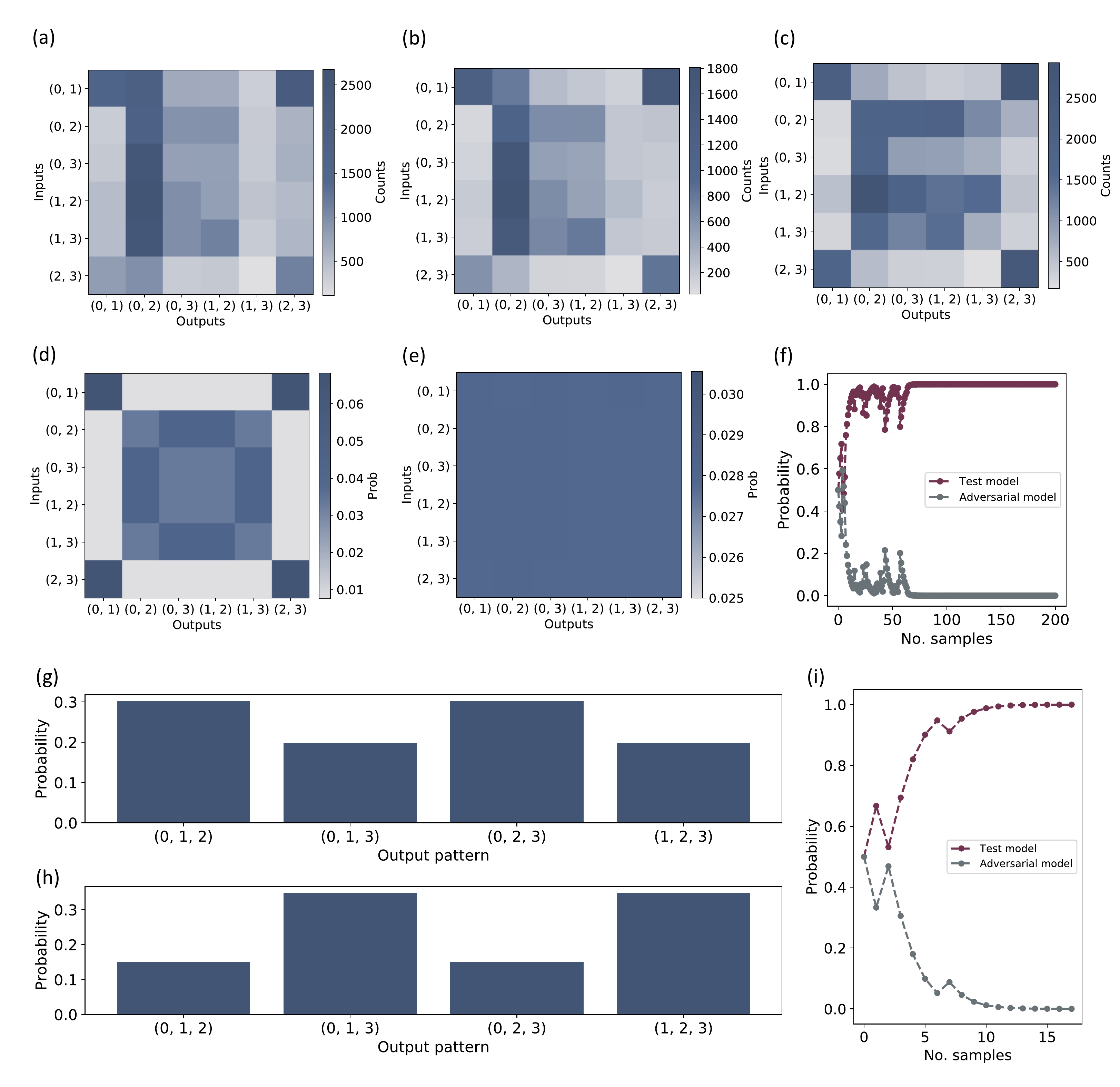}
    \caption{Boson sampling characterisation. (a) - (e) show the results of background subtraction and loss normalisation. (a) Raw measured counts. (b) Background subtracted counts. (c) Background subtraction and loss normalisation. (d) Simulated distribution for test model of indistinguishable photons plus mixedness. (e) Simulated distribution for adversarial model of non-interfering photons. (f) Bayesian verification for two-photon scattershot boson sampling. (g) Simulated distribution for three photon test model. (h) Simulated distribution for three photon adversarial model.  (i) Bayesian verification for three-photon boson sampling. }
    \label{fig:bs_charac}
\end{figure}
For both cases we verify against a test model that is indistinguishable interfering photons. To combat the 0 probability events corresponding to suppressed patterns in the Fourier interferometer, which however will occur due to experimental imperfections, we include a noninterfering term. The test model therefore becomes $P_{test} = r\vert\mathrm{Perm}\left(T_s\right)\vert^2 + (1-r)\mathrm{Perm}\left(\vert T_s \vert ^2\right)$, where $r$ parameterises the level of interference. We select $r$ to correspond to a pairwise overlap of 0.8; this value is also used in the main text boson sampling experiments. We use background subtraction techniques as demonstrated above and also normalise by the Klyshko efficiency of each output and the squeezing of the involved sources. For the two-photon interference case, the interferometer is set to implement a $F_4$ matrix with $\phi = \pi/2$. We note here that due to crosstalk between heaters, there is an offset to the value of $\phi$, leading to $\phi = $\SI{1.73}{\radian}. Operating in a scattershot configuration, we pump all four ring resonator sources and use two heralding photons to inform the input pattern and two signal photons to inform the output pattern. The probability is then calculated from the corresponding submatrix of $F_4$. Figure~\ref{fig:bs_charac}(d) shows the input/output probability distribution for this matrix with indistinguishable photons including mixedness which operates as our test model. Figure~\ref{fig:bs_charac}(e) shows the corresponding distribution for our adversarial model of noninterfering photons. In Fig.~\ref{fig:bs_charac}(a) we show the raw measured counts. Figures.~\ref{fig:bs_charac}(b) and (c) show the effects of background subtraction and both background subtraction and Klyshko rescaling, respectively. Figure.~\ref{fig:bs_charac}(f) shows dynamic Bayesian verification demonstrating that we are likely sampling from our test distribution. For the three-photon case the distribution of $F_4$ for both indistinguishable and distinguishable photons is uniform. To break this degeneracy, we tune the input MZIs. We choose the value empirically to be a value that minimises the indistinguishable/distinguishable distribution overlap but maximises the non-bunching probabilities to increase the overall 6-fold rate. Figures~\ref{fig:bs_charac}(g) and (h) show the distributions with both input MZI phases set to 0.9 for the indistinguishable and distinguishable cases, respectively. Figure~\ref{fig:bs_charac} shows the Bayesian verification again showing we are sampling from the desired distribution.
\subsection{Errors}\label{supp:errors}

In this section we state some relevant properties with their corresponding errors.
\begin{table}[h!]
    \centering
\begin{tabular}{|c|c|c|c|c|c|c|}
 \hline
  Experiment   & R1 location (nm) & R2 location (nm)  & R3 location (nm) & R3 location (nm) & $\Delta$(pm) & $\Delta$ (GHz)\\
  \hline
   HOM  & 1541.261$\pm$0.005 &1541.315$\pm$0.002& N/A&N/A&54.2 $\pm$ 5.2& 6.8 $\pm$ 0.7 \\
   Fusion  & 1541.263$\pm$0.005 &N/A& 1541.316$\pm$0.008&N/A&53.5 $\pm$ 9.5& 6.8 $\pm$ 1.2\\
   2-photon scattershot & 1541.26$\pm$0.015&1541.316$\pm$0.005&1541.312$\pm$0.005&1541.259 $\pm$ 0.03& (R1,R2) 54.8 $\pm$ 16.6 &  6.9 $\pm$ 2.1\\
   &&&&&(R1,R3) 50.5 $\pm$ 16.5 &  6.4 $\pm$ 2.1 \\
   &&&&&(R1,R4)  2.1 $\pm$ 16.3 &  0.3 $\pm$ 2.1  \\
   &&&&&(R2,R3) 4.1 $\pm$ 6.6 &  0.5 $\pm$ 0.8  \\
   &&&&&(R2,R4) 57.1 $\pm$ 5.7 &  7.2 $\pm$ 0.7  \\
   &&&&&(R3,R4) 52.9 $\pm$ 5.9 &  6.7 $\pm$ 0.7  \\
   3-photon boson sampling  & 1541.229$\pm$0.008&1541.277$\pm$0.006&N/A&1541.23$\pm$0.01&(R1,R2) 47.8 $\pm$ 10 &  6.04 $\pm$ 1.26\\
   &&&&&(R1,R4) 1.15 $\pm$ 14 &  0.1 $\pm$ 1.7 \\
   &&&&&(R2,R4) 46.7 $\pm$ 12 &  5.9 $\pm$ 1.5  \\
   
   \hline
\end{tabular}
    \caption[Ring locations]{Table showing the measured ring positions and corresponding detunings, $\Delta$, for the experiments in the main text} \label{tab:detuning}
\end{table}
In table~\ref{tab:detuning} we show the position of the ring resonators and detunings for each of the experiments in the main text. We measure the position of the ring resonator pump resonances by scanning the CW across the resonance and measuring the transmitted power. This is performed every two hours throughout each experiment and the values stated in table~\ref{tab:detuning} are the average and standard deviation of these positions. The average detuning and corresponding error are determined with Monte-Carlo sampling.

Table~\ref{tab:vis} shows the visibilities for different timing windows for both the HOM and fusion experiments. These errors are determined from the error on a sinusoidal fit of the fringe, including Poissonian error bars. 
For the fusion gate we also include the corresponding gate fidelity with accompanying errors calculated using Monte-Carlo sampling.

\begin{table}[h!]
    \centering
\begin{tabular}{|c|c|c|c|c|}
 \hline
  Experiment   & $\delta \tau = 200$  &  $\delta \tau = 100$ & $\delta \tau = 50$ &  $\delta \tau = 20$ \\
  \hline
   HOM &0.49 $\pm$ 0.01 &0.53 $\pm$ 0.01&0.58 $\pm$ 0.01&0.61 $\pm$ 0.01 \\
   Fusion  &0.40 $\pm$ 0.03 (0.79 $\pm$ 0.02)&0.46 $\pm$ 0.02(0.81 $\pm$ 0.01)& 0.56 $\pm$ 0.01 (0.859 $\pm$ 0.004) & 0.60 $\pm$ 0.01 (0.875 $\pm$ 0.004)\\
   
   \hline
\end{tabular}
    \caption{Table showing the visibilities for different timing windows for both HOM and fusion experiments. For the fusion gate, the corresponding gate fidelity is shown in parentheses} \label{tab:vis}
\end{table}

\end{document}